\RequirePackage{ifpdf}
\ifpdf 
\documentclass[pdftex]{sigma}
\else
\documentclass{sigma}
\fi

\begin{document}
\allowdisplaybreaks

\renewcommand{\PaperNumber}{025}

\FirstPageHeading

\ShortArticleName{Ordered Dissipative Structures in Exciton
Systems in Semiconductor Quantum Wells}

\ArticleName{Ordered Dissipative Structures in Exciton Systems\\
in Semiconductor Quantum Wells}

\Author{Andrey A. CHERNYUK and Volodymyr I. SUGAKOV}

\AuthorNameForHeading{A.A. Chernyuk and V.I. Sugakov}
\Address{Institute for Nuclear Research of NAS of Ukraine, 47
Nauky Ave., Kyiv, 03680 Ukraine}

\Email{\href{mailto:sugakov@kinr.kiev.ua}{sugakov@kinr.kiev.ua}}

\ArticleDates{Received November 10, 2005, in f\/inal form February
08, 2006; Published online February 23, 2006}

\Abstract{A phenomenological theory of exciton condensation in
conditions of inhomogeneous excitation is proposed. The theory is
applied to the study of the development of an exciton luminescence
ring and the ring fragmentation at macroscopical distances from
the central excitation spot in coupled quantum wells. The
transition between the fragmented and the continuous ring is
considered. With assumption of a defect in the structure, a
possibility of a localized island of the condensed phase in a
f\/ixed position is shown. Exciton density distribution is also
analyzed in the case of two spatially separated spots of the laser
excitation.}

\Keywords{dissipative structures; exciton condensation; quantum
wells} \Classification{82C22; 82D37}

\section{Introduction}
Indirect excitons in double quantum well (QW) structures consist
of electrons and holes driven to the separate wells by an external
electric f\/ield. They are long living particles due to the
hampered electron-hole (e-h) recombination. The enhanced lifetime
(more than 3 orders of magnitude longer than that of the direct
excitons in single QWs) makes it  possible to create large
concentrations of excitons in order to study ef\/fects of the
exciton-exciton interaction, an exciton condensation
\cite{1,2,3,4,5,6,7}, in particular.

In  AlGaAs and InGaAs based structures, a  non-trivial feature in
the photoluminescence spectra of excitons was observed
\cite{2,3,4,5,6,7}. A spot of the laser excitation was reported to
be surrounded by a concentric bright ring separated from the laser
spot by an annular dark intermediate region. The distance between
the ring and the spot grew with  increasing intensity of the
pumping and reached up to hundreds $\mu$\,m exceeding by much the
exciton dif\/fusion path. In certain cases, an internal ring was
observed nearby the laser spot.

At low temperatures (about 2 K), the external ring fragmented into
a structure with a~strongly evaluated periodicity on a macroscopic
scale \cite{2,4,6}. Fragments followed the external ring, even
when the excitation spot moved on the sample area or when the ring
radius varied with irradiation intensity. With temperature rising,
the external ring disappeared.

Photoluminescence intensity of indirect excitons strongly grew in
certain f\/ixed spatial regions~\cite{2,4}. For any allocation of
the excitation spot and its intensity, ``localized bright spots''
were observed, only when they were inside the area restricted by
the external ring. These spots washed out with temperature growth.

Under the excitation of the crystal by two spatially separated
laser spots, it was revealed that in the process of rapprochement,
the rings became deformed and slightly opened in the mutual
direction and f\/inally formed a common oval-like ring~\cite{4}.

The mechanism of the formation of the ring  was suggested in
\cite{6,7}. It was  based on two assumptions:
\begin{enumerate}\vspace{-2mm}\itemsep=0pt
\item[1)] in dark the well is populated with a certain density of
electrons; \item[2)] holes are captured by the well with a larger
probability than  electrons.\vspace{-2mm}
\end{enumerate}

As a result, the irradiated structure develops two dif\/ferently
charged spatial regions. The laser spot and the region around are
charged positively  while the region far away remains negative. As
electrons and holes can recombine only where they meet, the sharp
luminescence ring forms on the boundary between the regions of the
opposite charges.

Assuming an attractive interaction between excitons, in the
framework of a statistical app\-roach papers \cite{8,9} suggested an
explanation of the fragmentation of the ring  by means of 
the~nucleation of spots of a condensed phase. Models of phase
transitions in a system of indirect excitons were studied in
\cite{10}. Finite lifetime of the excitons leads to the
limitations on the size of the condensed phase areas, similarly to
e-h droplets in bulk semiconductors. In a 2D system, these areas
assume a shape of disk-like islands. The islands appear in the
places of the highest exciton generation, i.e.\ on the ring. Due
to the interaction between islands via the exciton concentration
f\/ields, there is a correlation in their positions eventually
resulting in a periodicity.

The statistical approach suggested in \cite{8,9} was based on the
following assumptions:
\begin{enumerate}\vspace{-2mm}\itemsep=0pt
\item[1)] the exciton generation rate on the ring is a smooth
function of coordinates, and is almost constant within the region
of a condensed phase island;

\item[2)] the distance between islands is much greater than the
size of an individual island;

\item[3)] the boundary between the gas and the condensed phase is
sharp.\vspace{-2mm}
\end{enumerate}

In the present work we go beyond the stochastic approach of
\cite{8,9} and remove the above assumptions and solve the problem
phenomenologically. We have incorporated a term containing free
energy of
 excitons into the equation controlling the  exciton density. The
resulting equation was solved with account of the pumping and
f\/inite lifetime of excitons. This approximation allows
treatment of inhomogeneous systems and obtaining new results, for
instance, description of transition from a fragmented to a
continuous luminescence ring with changing parameters of the
system,  development of an internal ring in additions to the
internal one, an explanation of localized spots in the emission, a
deviation of the shape of fragments from the
 spherical, the excitation of the system by two laser spots. Study of
these essentially inhomogeneous problems in the framework of the
statistical model would be very complicated.

A mechanism leading to  instability in exciton subsystem was
suggested in the paper \cite{11}, yet the problem of the
structures of the exciton density forms outside the laser spot at
a non-uniform pumping was not considered.

\section{The exciton generation rate on the ring}
In order to determine the exciton generation rate, we use the
model of the ring formation suggested in \cite{6,7}. We shall
obtain solutions for the cases, which will be applied in the
further study of the fragmentation. The equations for the electron
and hole densities in a QW plane outside the laser spot are
\begin{gather}\label{eq1}\frac{\partial n_e}{\partial t}=D_e \Delta n_e+K_e-Wn_e n_h-\frac{n_e-n_e^0}{\tau_e},
\\
\label{eq2}\frac{\partial n_h}{\partial t}=D_h \Delta n_h+K_h-Wn_e
n_h-\frac{n_h}{\tau_h}.
\end{gather}
$n_e$ ($n_h$) is the electron (hole) concentration,
 $D_e$ ($D_h$) is the electron (hole) dif\/fusion coef\/f\/icient,
$\tau_e$ ($\tau_h$) characterizes the time of the establishment of
the electron (hole) equilibrium  between the well and the crystal
outside, $W$ is the e-h recombination rate, $n_e^0$ is the
electron concentration in the well in  absence of the irradiation.
$K_e$ ($K_h$) is the electron (hole) creation rate, the radial
prof\/ile of which is approximated by a Gaussian curve
\begin{gather}\label{eq20}
K_{e,h}\left(\rho\right)=C_{e,h}\frac{1}{l \sqrt{2\pi}}
\exp\left({-\frac{\rho^2}{2l^2}}\right).
\end{gather}
The exciton generation rate is proportional to the product of
$n_e$ and $n_h$:
\begin{gather}\label{eq4}
G=qWn_e n_h,
\end{gather}
where $q$ is the share of recombined electrons and holes, which
participate in exciton formation ($q<1$).

The system of the equations~(\ref{eq1}) and (\ref{eq2}) was solved
numerically on a disk or on a rectangular plate with the sizes
considerably exceeding the ring radius. The task was solved with
the following initial conditions: $n_e({\boldsymbol r})=n_e^0$,
$n_h({\boldsymbol r})=0$ at $t=0$, and the boundary conditions:
$n_e\left(t\right)=n_e^0$, $n_h\left(t\right)=0$ at
$r\rightarrow\infty$. We chose the following parameters of the
system: $D_e=200\,{\rm cm}^2\,{\rm s}^{-1}$, $D_h=50\,{\rm
cm}^2\,{\rm s}^{-1}$, $\tau_e=\tau_h=10^{-5}$\,s,
$n_e^0=10^{-10}$\,cm$^{-2}$, $l=$60$\mu$\,m, $q=0.9$. The values
of $D_e$, $D_h$, $\tau_e$, $\tau_h$ coincide with the values in
\cite{5}.

The Fig.~1 shows the steady-state solutions of  the
equations~(\ref{eq1}) and (\ref{eq2}). Radial distribution of the
exciton density  has a sharp maximum at a certain distance from
the center. This maximum moves farther away from the laser spot
with increasing the pumping (see curves 1, 2 in Fig.~1).

\begin{figure}[t]
\centerline{\includegraphics[width=9cm]{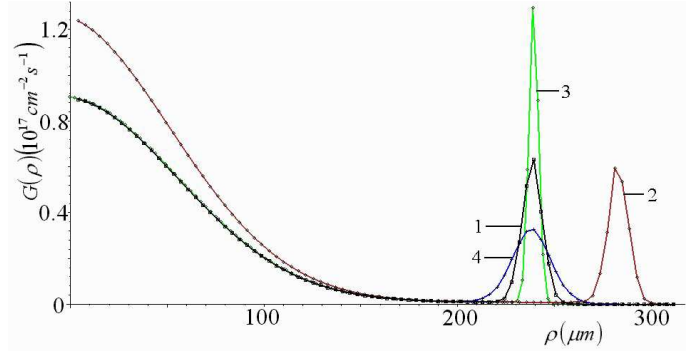}}

\vspace{-3mm}

\caption{The radial prof\/ile of the exciton generation rate
$G(\rho)$ in a QW plane.
 Pumpings are $C_e=1.5\cdot 10^{15}\,{\rm cm}^{-1}\,{\rm s}^{-1}$,
  $C_h=3\cdot 10^{15}\, {\rm cm}^{-1}\, {\rm s}^{-1}$ for the curves 1, 3, 4 and 1.33 times larger for
 the curve 2. $W$~is 3\,cm$^2\, {\rm s}^{-1}$ for the curves 1 and 2;
 22\,cm$^2\,{\rm s}^{-1}$ for the curve 3; 0.8\,cm$^2\,{\rm s}^{-1}$ for the curve 4.}
 \end{figure}

Width of the peak depends on the e-h recombination rate $W$. This
rate is determined by the tunneling processes and depends on the
width of the barrier which separates the QWs in the double well
structure. With increasing $W$, the peak does not change its
position, but narrows down and grows in height (compare curves 1,
3, 4 in Fig.~1). Its integral
 value remains approximately, within several percent, constant.

\section{Model of the system. Exciton density equation}

The microscopic theory of  exciton condensation at inhomogeneous
excitation is a dif\/f\/icult task, because we must take into
account relaxation and  f\/inite exciton lifetime. Therefore, to
build a solution, we apply a phenomenological approach. Our
treatment of the problem is based on the following assumption: we
assume that  time of the post-excitation establishment of a
quasi-local equilibrium of electron and hole densities and their
binding into excitons is short, much smaller than  time of the
equilibration between dif\/ferent areas. The latter is related to
slower dif\/fusion processes. In this case,
 free energy of the quasi-equilibrium state can be considered as a function of the
exciton density, which depends on the spatial coordinates.
Obviously, such  description is possible at macroscopic distances,
much larger than the exciton free path, on which the wave function
loses coherence. Then, a phenomenological equation for the exciton
density can be written down as
\begin{gather}\label{eq5}
\frac{\partial\,n_{\rm ex}}{\partial\,t}=- \mbox{div}\,
{\boldsymbol j} +G-\frac{n_{\rm ex}}{\tau_{\rm ex}},
\end{gather}
where $n_{ex}$ is the exciton density, $\tau_{\rm ex} $ is the
exciton lifetime, ${\boldsymbol j}$ is the density of the exciton current:
\[
{\boldsymbol j}=-M \nabla\mu,
\]
$M$ is the exciton mobility:
\[
M=\frac{n_{\rm ex}^{c}D_{\rm ex}}{k_{B}T},
\]
$\mu$ is the chemical potential:
\[
\mu=\frac{\delta\,F}{\delta\,n_{\rm ex}}.
\]
We chose  free energy in the form suggested by the Landau model:
\begin{gather}\label{eq6}
F[n_{\rm ex}]=\int{d {\boldsymbol r}\left[{\frac{K}{2}\left({\nabla n_{\rm
ex}}\right)^2+f\left({n_{\rm ex}}\right)}\right]}.
\end{gather}
Here, the term $K\left({\nabla n_{\rm ex}}\right)^2/2$
characterizes the energy of an inhomogeneity. Free energy density~$f$ is
\begin{gather}\label{eq7}
f\left({n_{\rm ex}}\right)=\frac{a}{2}\left({n_{\rm ex}-n_{\rm
ex}^c}\right)^2 +\frac{c}{3}\left({n_{\rm ex}-n_{\rm
ex}^c}\right)^3+\frac{b}{4} \left({n_{\rm ex}-n_{\rm ex}^c}
\right)^4,
\end{gather} $a$, $b$, $c$, $n_{\rm ex}^c$ are
phenomenological parameters, which can be obtained from the
quantum mechanical calculations, or extracted from the comparison
of the theory with  experiment.

After substituting equations~(\ref{eq6}) and (\ref{eq7}) into
equation~(\ref{eq5}), the latter becomes
\begin{gather}\label{eq8}
\frac{\partial\,n_{\rm ex}}{\partial\,t}=-\Delta^2n_{\rm
ex}+\left(3{n_{\rm ex}^c}^2-2\beta n_{\rm ex}^{c}-1\right) \Delta
n_{\rm ex}- \frac{n_{\rm ex}}{\tau_{ex}}+G-\left(3n_{\rm
ex}^c-\beta\right)\Delta {n_{\rm ex}}^2+\Delta{n_{\rm ex}}^3,
\\
\beta\equiv\frac{c}{\sqrt{-ab}}.\nonumber
\end{gather}
Here we introduce  dimensionless units
\[\xi=\sqrt{\frac{K}{-a}}, \qquad \tau=\frac{K}{M a^2}, \qquad
n^0_{\rm ex}=\sqrt{\frac{-a}{b}}
\]
for length, time and exciton concentration, respectively. We
assume that equation~(\ref{eq8}) holds also if the condensed phase
is an e-h liquid. Then $F$ is a function of the density of e-h
pairs.

The phenomenological equation~(\ref{eq8}) for  study of periodical
 structures in exciton systems was employed in~\cite{12} for the case of
 spatially uniform exciton generation  in bulk materials.
In the case of inf\/inite exciton lifetime (excitons are neither
created nor decay), the above equation describes equilibrium
states, i.e.\ two phases with the densities
\begin{gather}\label{eq9}
n_{\rm ex}^\pm=n_{\rm ex}^c\pm1
\end{gather}
(in dimensionless units). The solution $n_{\rm ex}^-$ determines
the  density of the gas phase, while $n_{\rm ex}^+$ stands for
the condensed one. In the non-equilibrium system with a f\/inite
value of the exciton lifetime $\tau_{\rm ex}$, the  stationary
uniform solutions of equation~(\ref{eq9}) are unstable. The
exciton density varies in space periodically  in the range limited
by certain values  $n_{\min}$ and $n_{\max}$. In the case of the
f\/inite lifetime $n_{\min}$ and $n_{\max}$ can dif\/fer
essentially from their equilibrium values.  In the limit of the
long exciton lifetimes $\left({\tau_{\rm ex}\to\infty}\right)$
$n_{\min}$ and $n_{\max}$ approach, respectively, their values in
the gas $n_{\rm ex}^-$ and the condensed $n_{\rm ex}^+$ phases.

In the given contribution the generation rate G is an essentially
inhomogeneous function of coordinates. The system of the
equations~(\ref{eq1}), (\ref{eq2}), (\ref{eq8}) is a closed system
of the equations, which enables describing  phase states in the
exciton system under inhomogeneous excitation.
\section{Numerical simulation of exciton condensed phase formation\\
 on the ring}

\subsection{Exciton concentration distribution}

Having determined the exciton generation rate by solving
equation~(\ref{eq4}), we numerically studied the stationary
solutions of the nonlinear Eq.~(\ref{eq8}) for the exciton
density. We chose the following parameters of the system: $D_{\rm
ex}=10\,{\rm cm}^2\,{\rm s}^{-1}$, $\tau_{\rm ex}=10^{-7}$\,s,
$\xi=4\mu$\,m, $\tau= 3\cdot 10^{-9}$\,s, $n^0_{\rm ex}=
1.875\cdot 10^9\,{\rm cm}^{-2}$, $n_{\rm ex}^c= 1.2n^0_{\rm ex}$,
$T=2$\,K, $\beta=0.015$. The simulation showed that for the
exciton generation rate of equation~(\ref{eq4}) in Fig.~1, the
excitons condense  into a ring which can break down into separate
fragments (see Fig.~2a). There were 28 and 36 islands on the rings
of the radii $242\mu$\,m and $295\mu$\,m for the curves 1 and 2 of
Fig.~1, respectively. An island shape is oval (of the order of
$20\mu\,{\rm m}\times 30 \mu$\,m at the half-maximum), which is
determined by the parameter~$K$.

Strictly speaking, in the system with strongly non-uniform density
distribution, it is impossible to assign certain phase states to
separate areas, nevertheless points of high density can be
attributed to the condensed phase, and points of low density to
the gas phase. Thus, the created fragments are periodically
positioned islands of the condensed phase of excitons, emergence
of which is caused by the f\/inite value of lifetime and the
inter-exciton attraction. The condensed phase comprised by a
periodic array of exciton islands corresponds to the fragments
observed in experiments \cite{2,4,6}.

\begin{figure}[t]\centerline{\includegraphics[width=7.7cm]{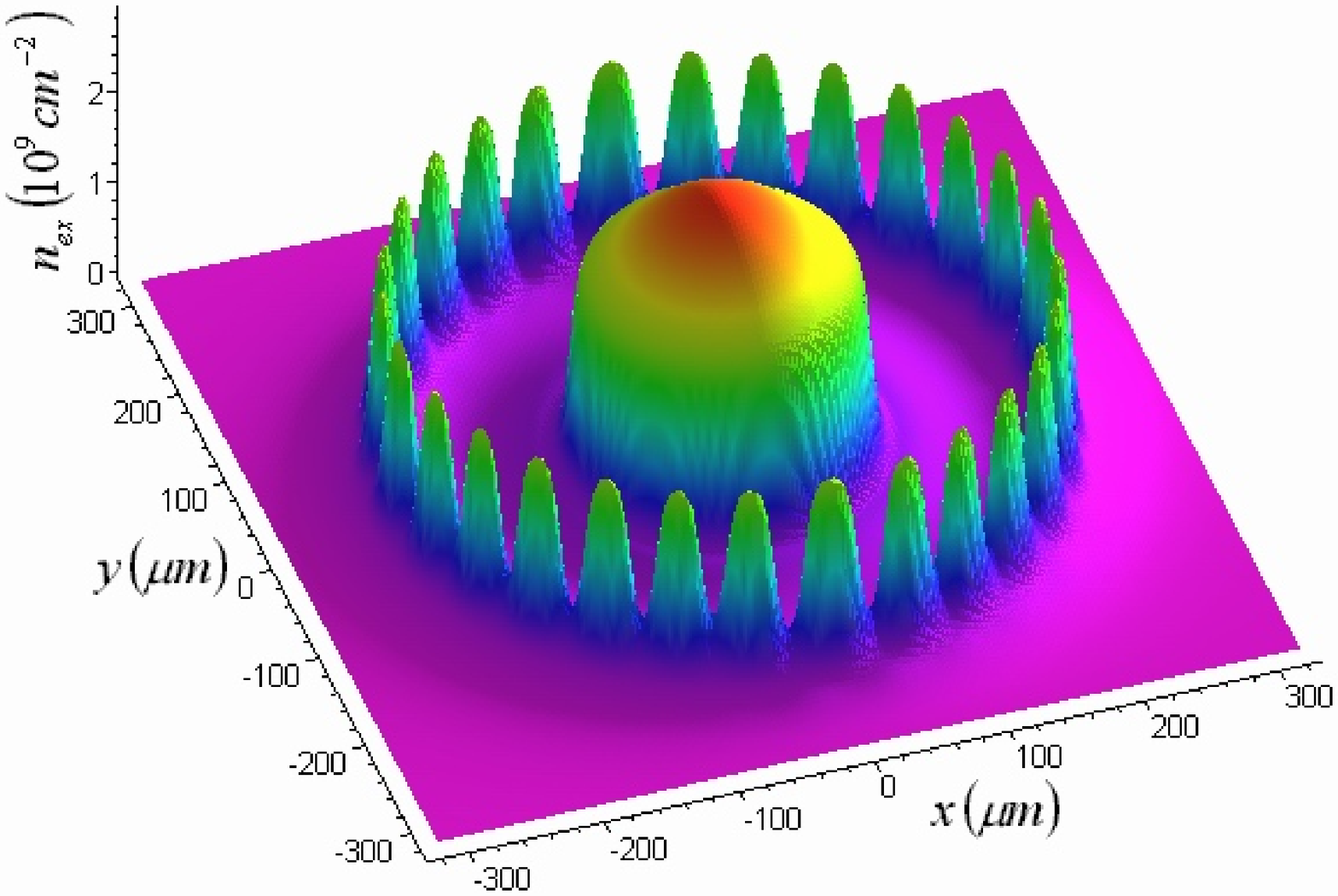}\hfil
\includegraphics[width=7.7cm]{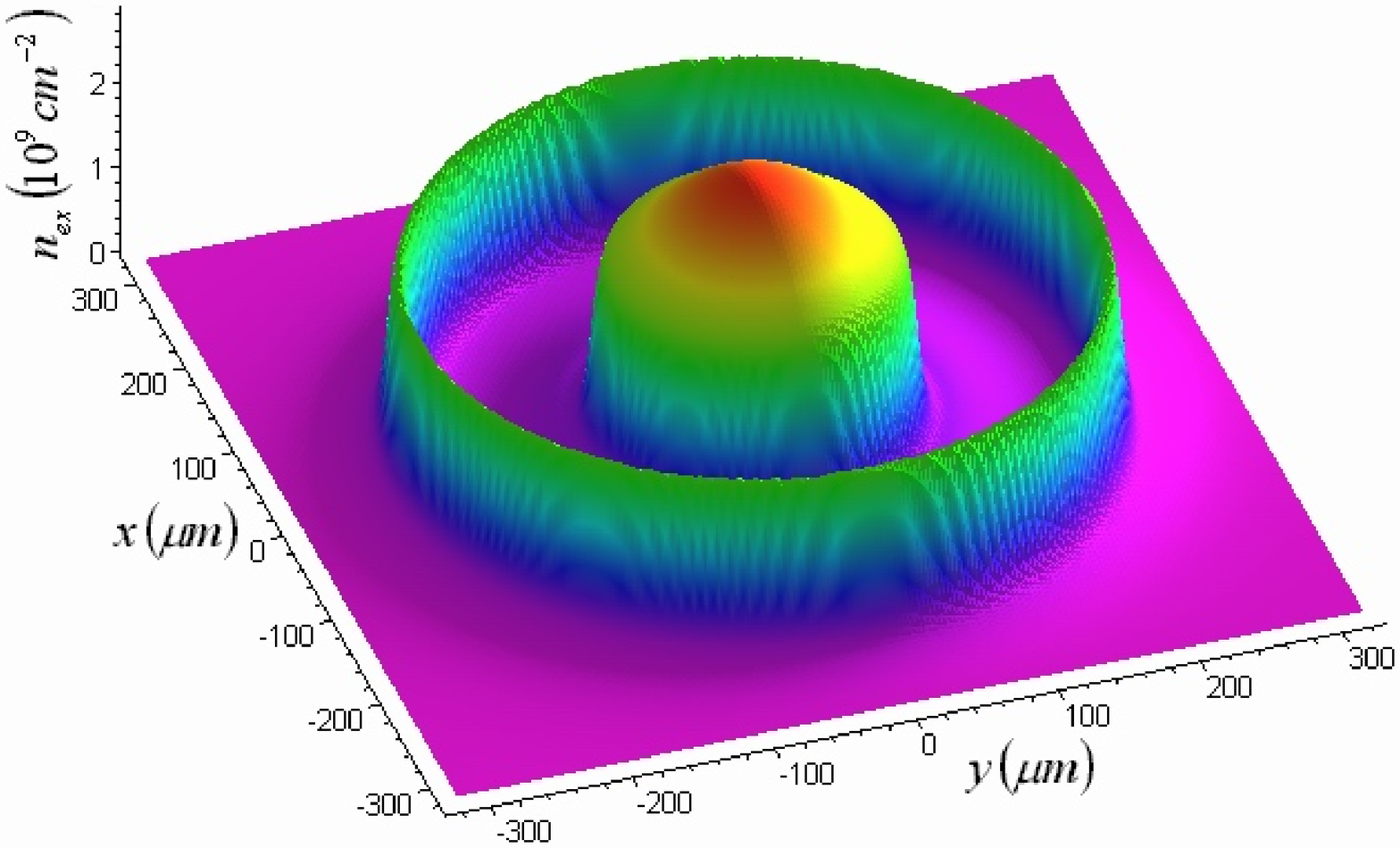}}
\vspace{-3mm}

\caption{Exciton density $n_{\rm ex}({x,y})$ for the parameters
corresponding to the generation rate given in Fig.~1: a) for the
curve 1, b) for the curve 3.}
\end{figure}

The radius of the ring of the condensed phase grows in two cases:

\begin{enumerate}\itemsep=0pt\vspace{-2mm}
 \item[1)] with increasing the pumping;

\item[2)] if the laser spot expands leaving the integral intensity
constant.\vspace{-2mm}
\end{enumerate}

With expanding the ring radius, the number of the exciton islands
increases. They remain of the same size, and the distance between
them  is practically invariant. Similar trends such as  increased
radius of the exciton ring with increasing the irradiation
intensity and in the case of the expansion of the excitation spot
were observed in \cite{2,3,4,5,6,7}.

Some experiments \cite{2,4,6} report  observations of an inner
ring in the photoluminescence emission. We attribute   emergence
of the internal ring  to specif\/ic character of the irradiation
intensity prof\/ile in the vicinity of the laser spot edge. If the
exciton generation near the laser spot does not change abruptly
then,  as our study shows, an internal concentric ring may appear
around the central maximum. Our calculations show that it can be
fragmented as well. Yet, we have to note that  application of our
theory is limited by  assumption that the exciton gas is believed
to be cooled, which might not be true near the laser spot.
Possible reasons of the inner ring appearance due to the
non-homogeneous exciton distribution and temperature are stated in
the paper \cite{13}. Additional investigations at dif\/ferent
pumpings are needed for the f\/inal clarif\/ication of the inner
ring formation.

As  already specif\/ied, with  increasing  the e-h recombination
rate $W$  exciton generation rate becomes sharper (compare the
curves 1 and 3 in Fig.~1). The results of the calculations show
that with increasing $W$ the fragmented ring of the condensed
phase becomes continuous (see Fig.~2b). This is seen from the
comparison of Figs.~2a and 2b, obtained for the dif\/ferent values
of $W$ ($3\,{\rm cm}^2\,{\rm s}^{-1}$ and $22\, {\rm cm}^2\, {\rm
s}^{-1}$, respectively) and for the same values of other
parameters.

With decreasing $K$, the island size of condensed excitons
diminishes, so the fragmented ring  passes to continuous one.
Increase of $K$ gives rise to continuous ring formation, because
islands grow. The fragments do not form because the surface energy
is smaller in the case of the cylindrical shape of a fragment. As
$K$ is related with the surface energy, the role of the surface
energy with $K$ changing is not def\/ining and the solid circle
forms in the system.

The equation~(\ref{eq8}) can have solutions both as a continuous
ring and as periodically placed islands of condensed excitons, and
the choice of one of them can depend on the inhomogeneity of the
system or on the boundary (initial) conditions.

\subsection{Localized spots}
The appearance of the ``localized bright spots'' can be attributed
to non-uniformities of dif\/ferent nature in the QW \cite{2,4}. We
have undertaken an attempt to simulate them in our approach,
assuming that there is a certain region that dif\/fers from the
surrounding regions by quality of the contact between the well and
the matrix in which the wells are grown.  We assumed that the
value of $\tau_e$, which characterizes the time of the
establishment of the electron density equilibrium between the
wells and the matrix, varies in  space and expressed this
non-uniformity in the vicinity of the point ${\boldsymbol
r}_0=(L,0)$ as
\begin{gather}\label{eq10}
\tau_e({\boldsymbol r})=\frac{\tau_e}{1+B\exp\left[
{-\frac{\left({x-L}\right)^2+y^2}{2c^2}}\right]}.
\end{gather}
We carried out  calculations of the exciton density  prof\/ile
governed by equation~(\ref{eq8}). An additional island located in
the region of the non-uniformity emerges in this case (see
Fig.~3). Obviously, in the case of several defects, several
islands would appear as observed in \cite{2,4,6}. This is 
a~consequence of non-linearity of the system. So, inhomogeneity,
which arises in the crystal as a~result of technological
processes, can be a nucleus of new fragments.

\begin{figure}[t]
\centerline{\includegraphics[width=8cm]{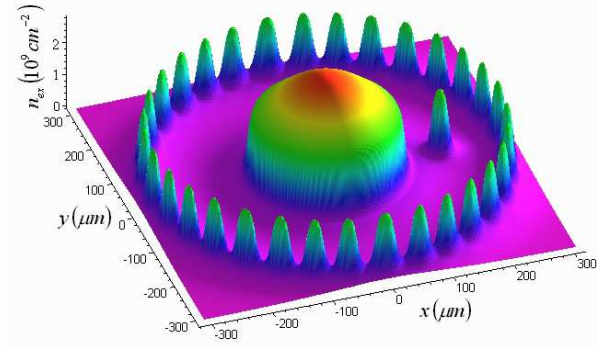}}

\vspace{-3mm}

\caption{Emergence of a localized spot in the exciton density
distribution at a
 point of local non-uniformity given by equation~(\ref{eq10}) with parameters
 $B=150$, $L=190\mu$\,m, $c=12.2 \mu$\,m.}
 \end{figure}

 Similar calculations can be performed for  inhomogeneities
 of a different nature. So, the presented theory can  describe  condensed
 states localized near defects.

\subsection{Modelling of two spots excitation}
We also carried out a simulation of the excitation of the system
by two irradiation sources for  description of the experiment with
two spatially separated lasers \cite{3}. We solved the
equations~(\ref{eq1}) and (\ref{eq2}) with two laser sources with
the radial shape (\ref{eq20}) each. Our calculations show
non-linear contributions of the excitation from two spots: rings
become extended in  mutual directions and at a certain stage of
spots  approaching merge into a common oval-shaped ring (see
Fig.~4). So, in such a way the rings can interact.

\begin{figure}[t]

\centerline{\includegraphics[width=7.5cm]{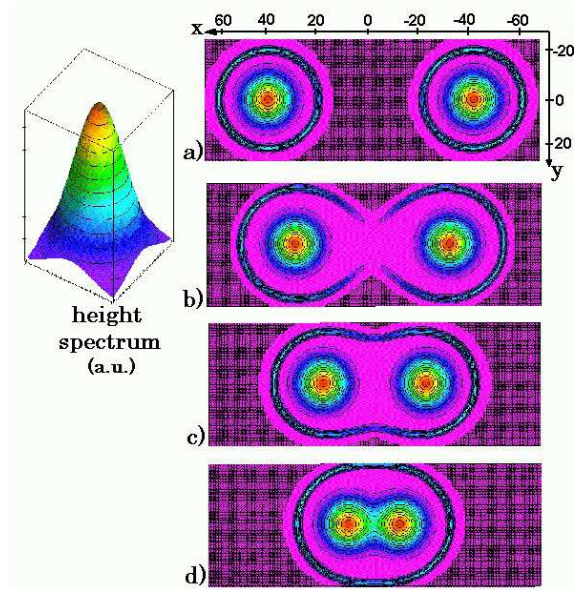}}
\vspace{-3mm}

\caption{Exciton density distribution in the system irradiated by
two laser spots. The
 distance between the spots centers are: a) 84$\mu$\,m, b) 61$\mu$\,m, c) 42$\mu$\,m, d)
 21$\mu$\,m. The pumpings are smaller than in Fig.~2.}
 \end{figure}

The results in Fig.~4 conf\/irm the non-additive character of the
rings interaction, observed in the experiment~\cite{3}.

\subsection{Behavior versus temperature}

We studied \cite{3} behavior of the system versus temperature. We
assumed that the parameter $a$ (see equation~(\ref{eq7})) can be
represented according to the Landau model:
\[
a=\alpha\left({T_c-T}\right),
\]
where $\alpha<0$ and $T_c$ is the critical temperature. With
temperature rising, the fragmentation of the ring disappears, its
intensity  falls, the radial structure of the exciton density
becomes dim (see Fig.~5). Such transition with temperature growth
is observed also in \cite{2}.

\begin{figure}[t]
\centerline{\includegraphics[width=8cm]{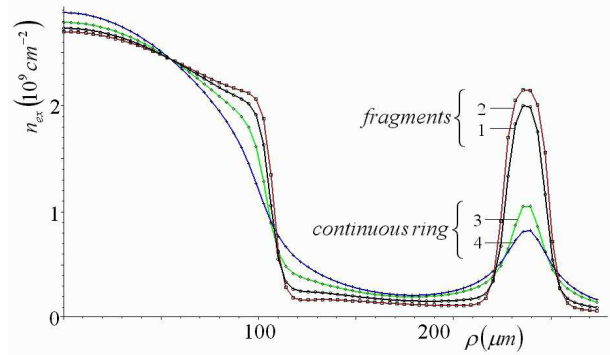}}

\vspace{-3mm}

\caption{The radial prof\/ile of the exciton density $n_{\rm
ex}(\rho)$ at dif\/ferent
 temperatures: 1) 2\,K, 2) 1.6\,K, 3)~2.7\,K, 4) 4.4\,K. $T_c=3.7$\,K.
 The curves 1 and 2 represent the fragmented ring, the curves~3 and 4 are the prof\/iles
 of low continuous rings.}
 \end{figure}

\section{Conclusions}
Thus, the results of our  study of the exciton condensation in a
coupled QW around the laser spot can be summarized  in  the
following conclusions:
\begin{enumerate}\vspace{-2mm}\itemsep=0pt
\item Fragmentation of the external ring of exciton density occurs
at a certain threshold value of the exciton generation rate
(pumping).

\item Transition from a fragmented ring to a continuous one occurs
at increased value of the electron-hole recombination rate (i.e.\
for a narrower barrier between the QWs), with reduction of the
surface energy (the parameter $K$) and at higher temperature.

\item In a QW with a macroscopic defect a localized island of the
condensed phase may emerge inside the external ring.

\item Possibility of the development of an internal ring depends
on the character of decrease of irradiation intensity at the edge
of the
 laser spot.

\item Exciton luminescence rings caused by two spatially separated
laser spots attract and eventually form a common oval-shaped ring.
\end{enumerate}

The considered periodical structures exist only due to the
external excitation and are the development of self-organization
in the system of exciton condensed phases. According to the
classif\/ication of \cite{14}, structures of such type which exist
exclusively in non-equilibrium conditions are called dissipative.

\subsection*{Acknowledgements}
The work was supported by INTAS grant No.03-51-5266 and by the
Ukrainian Ministry of Education and Science (project
No.02.07/147).
 The authors thank to Dr.~I.Yu.~Goliney for useful discussions.

\LastPageEnding

\end{document}